# Growth and characterization of n-type electron-induced ferromagnetic semiconductor (In,Fe)As

Pham Nam Hai[1], Le Duc Anh[1], Shyam Mohan[2], Tsuyoshi Tamegai[2], Masaya Kodzuka[3], Tadakatsu Ohkubo[4], Kazuhiro Hono[3,4], and Masaaki Tanaka[1,5*],

1. Department of Electrical Engineering and Information Systems, The University of Tokyo, 7-3-1 Hongo, Bunkyo-ku, Tokyo 113-8656, Japan

2. Department of Applied Physics, The University of Tokyo, 7-3-1 Hongo, Bunkyo-ku, Tokyo 113-8656, Japan

3. Graduate School of Pure and Applied Sciences, University of Tsukuba, Tsukuba 305-0047, Japan

4. National Institute for Materials Science, 1-2-1 Sengen, Tsukuba 305-0047, Japan

5. Japan Science and Technology Agency, 4-1-8 Honcho, Kawaguchi-shi, Saitama 332-0012, Japan

We show that by introducing isoelectronic iron (Fe) magnetic impurities and Beryllium (Be) double-donor atoms into InAs, it is possible to grow a n-type ferromagnetic semiconductor (FMS) with the ability to control ferromagnetism by both Fe and independent carrier doping by low-temperature molecular-beam epitaxy. We demonstrate that (In,Fe)As doped with electrons behaves as an n-type electron-induced FMS. This achievement opens the way to realize novel spin-devices such as spin light-emitting diodes or spin field-effect transistors, as well as helps understand the mechanism of carrier-mediated ferromagnetism in FMSs.

In the emerging field of "semiconductor spintronics", carrier-induced ferromagnetism in ferromagnetic semiconductors (FMSs) is the central issue. For two decades, most studies on FMSs have been concentrated on III-V semiconductors with tetrahedral Mn-As bonding, such as (In,Mn)As [1] or (Ga,Mn)As [2,3], which are always p-type with hole densities as high as $10^{20}$ - $10^{21}$ cm$^{-3}$. In those materials, Mn atoms work not only as local magnetic moments but also as acceptors providing holes that mediate ferromagnetism. For practical applications to electronic devices, not only p-type but also n-type carrier-induced FMSs are required. In this letter, we show that by introducing isoelectronic Fe magnetic impurities and Be double-donor atoms into InAs, it is possible to grow a n-type FMS with tetrahedral Fe-As bonding by low-temperature molecular-beam epitaxy (LT-MBE). We demonstrate that (In,Fe)As doped with electrons behaves as an n-type electron-induced FMS.

The studied 100 nm-thick $(In_{1-x},Fe_x)As$ layers were grown by LT-MBE on semi-insulating GaAs substrates as follows. After growing a 50 nm-thick GaAs buffer layer at 580°C, we grew a 10 ~ 20 nm-thick InAs buffer layer at 500°C. The growth of InAs at high temperature helps to quickly relax the lattice mismatch between InAs and GaAs, and create a smooth InAs surface. After cooling the sample down to 236°C, we started growing a 100 nm–thick (In,Fe)As layer with or without Be co-doping. Finally, we grew a 5 ~ 20 nm InAs cap layer to prevent oxidation of the underlying (In,Fe)As layer. Two series of $(In_{1-x},Fe_x)As$ samples were grown as summarized in Table I. Series A with a Fe concentration of $x$ = 5.0% and series B with a higher Fe concentration of $x = 8.0\%$ (except for B0 with $x = 9.1\%$) were grown at a substrate temperature of 236°C, with and without electron doping. Figure 1(a) shows a transmission electron microscopy (TEM) image of sample B0, which is $(In_{0.909},Fe_{0.091})As$ without Be co-doping. Figure 1(c) shows a high-resolution TEM lattice image of an area close to the buffer layer, indicated by the red rectangle in Fig. 1(a). The inset in Fig. 1(c) shows a transmission electron diffraction (TED) pattern of this (In,Fe)As layer. It is clearly seen from the TEM and TED that the whole (In,Fe)As layer is zinc-blende crystal structure and there is no visible metallic Fe or inter-metallic Fe-As precipitation, which is also confirmed by X-ray diffraction. Figure 1(b) shows In, Fe and As atomic concentrations obtained by energy dispersive x-ray (EDX) spectroscopy in sample B0. It is observed that the As atomic concentration is close to the sum of the In and Fe atomic concentrations, revealing that Fe atoms reside at the In sites. Furthermore, we have used the laser-assisted three-dimensional atom probe (3DAP)[4] to investigate the distribution of Fe, In, and As in (In,Fe)As with a nearly atomic resolution. Figure 1(d) shows the three-dimensional atom distributions of In, Fe and As obtained by 3DAP for sample A4, which is Be-doped $(In_{0.905},Fe_{0.05})As$. It can be seen that the Fe atoms distribute everywhere in the (In,Fe)As layer. In order to find any precipitation of metallic Fe or inter-metallic Fe-As nanoclusters, we divided all of the observed area into about 18,000 blocks (200 atoms / block) and investigated the local Fe, In and As distributions in those blocks. Atomic distributions of In, Fe and As in all blocks can be described exactly by the chemical formula of (In,Fe)As, with no block containing metallic Fe nanoclusters (Fe 100%, In 0%, As 0%) or inter-metallic Fe-As nanoclusters (In 0%). All of these results confirmed that our (In,Fe)As samples are composed of a single-phase zinc-blende crystal.

At the In sites, the Fe ions have two possible states; acceptor state ($Fe^{2+}$) and neutral state ($Fe^{3+}$). Experimentally, sample B0 (and all the other samples without Be doping, not shown here) shows n-type with a maximum residual electron concentration of $1.8\times10^{18}$ cm$^{-3}$ at room temperature, which is four orders of magnitude smaller than the doped Fe concentration. Our analysis of the temperature dependence of the electron mobility of sample B0 shows that the neutral impurity scattering, rather than the ionized impurity scattering, is the dominant scattering mechanism up to room temperature (see Fig. S1 in Supplementary Information). All of these facts indicate that the Fe atoms in (In,Fe)As are in the neutral state ($Fe^{3+}$) rather than the acceptor state ($Fe^{2+}$). This result is consistent with the chemical trend of Fe in III-V semiconductors [5], and the results of electron paramagnetic resonance of Fe impurity in InAs, which shows the isoelectronic $Fe^{3+}$ state with $3d^5$ configuration (5 $\mu_B$ / Fe atom) [6]. The residual electrons in sample B0 probably come from the As anti-site defects acting as donors due to the LT-MBE growth [7].

We then tried doping (In,Fe)As layers with donors while fixing the Fe concentration to see the carrier-induced ferromagnetism. After trying several doping methods, we found that Beryllium (Be) atoms doped in (In,Fe)As at a low growth temperature of $T_S$ = 236°C work as good double donors, not as acceptors as in Be-doped InAs grown at $T_S$ > 400°C. The n-type conduction of (In,Fe)As was confirmed by the normal Hall effect and thermoelectric Seebeck effect (see Supplementary Information). For these electron doped (In,Fe)As layers, we investigate their ferromagnetism by using magnetic circular dichroism (MCD) and superconducting quantum interference device (SQUID). Despite the general belief that the tetrahedral Fe-As bonding is antiferromagnetic [8], all of our data show evolution of ferromagnetism in (In,Fe)As with increasing the Fe concentration ($x$ = 5 - 8%) or with increasing the electron concentration ($n$ = 1.8×10$^{18}$ cm$^{-3}$ to 2.7×10$^{19}$ cm$^{-3}$), indicating that (In,Fe)As is an intrinsic n-type FMS, and that we can control the ferromagnetism of this material independently by Fe doping and electron doping.

MCD is a technique that measures the difference between the reflectivity for right ($R_{\sigma+}$) and left ($R_{\sigma-}$) circular polarizations: $\mathrm{MCD} = \frac{90}{\pi} \frac{(R_{\sigma+} - R_{\sigma-})}{2} \sim \frac{90}{\pi} \frac{dR}{dE} \Delta E$, where $R$ is the reflectivity, $E$ is the photon energy, and $\Delta E$ is the spin-splitting energy (Zeeman energy) of a material. Since the MCD spectrum of a FMS directly probes its spin-polarized band structure and its magnitude is proportional to the magnetization ($\Delta E \propto M$), MCD is a powerful and decisive tool to judge whether a FMS is intrinsic or not [9]. Note that the spectral features (i.e. enhanced at optical critical point energies of the host semiconductor), rather than the absolute magnitude of MCD, are important to judge whether a FMS is intrinsic or not. Figure 2 shows the MCD spectra of sample series A (A1 - A4) and sample series B (B1 - B4), measured at 10 K under a magnetic field of 1 Tesla applied perpendicular to the film plane. With increasing the electron density while fixing the Fe concentration, the MCD intensity shows strong enhancement at optical critical point energies $E_1$ (2.61 eV), $E_1 + \Delta_1$ (2.88 eV), $E_0'$ (4.39 eV) and $E_2$ (4.74 eV) of InAs, which show the magnetic "fingerprints" of (In,Fe)As. For sample B4, $(In_{0.92},Fe_{0.08})As$ with $n = 2.8 \times 10^{19}$, the MCD peak at $E_1$ already reaches 100 mdeg at 10 K, which is two orders of magnitude greater than the MCD caused by the Zeeman splitting of InAs (~1 mdeg/Tesla)[9]. Therefore, the effective magnetic field acting on the InAs matrix is as large as 100 Tesla; thus, it can not be explained by the stray field of embedded ferromagnetic Fe nanoclusters, if any. Furthermore, we show in Fig. 2(i) the MCD spectrum of a 44 nm-thick Fe thin film grown on a GaAs substrate at 30°C. The MCD spectra of our (In,Fe)As samples are clearly different from that of Fe, thus eliminating the possibility of metallic Fe particles. These results further confirm that (In,Fe)As maintains its zinc-blende structure, and that its band structure is spin-split. Samples A4, B3 and B4, whose electron concentrations are about $10^{19}$ cm$^{-3}$, are ferromagnetic, while other samples with lower electron concentrations are paramagnetic. These facts also indicate that the ferromagnetism of (In,Fe)As is induced by electrons and that we can rule out embedded metallic Fe or intermetallic Fe-As compound nanoparticles (if any) as the source of the observed ferromagnetism. Note that there are three intermetallic Fe-As compounds in their binary phase diagram: $FeAs_2$, FeAs and $Fe_2As$ [10].

However, none of them is ferromagnetic; FeAs$_2$ is diamagnetic [11], while FeAs and Fe$_2$As are both anti-ferromagnetic with Neel temperature of 77 K and ~ 353 K, respectively [10]. We also observed anomalous Hall effect (see Supplementary Information), anisotropic magnetoresistance and planar Hall effect in (In$_{0.94}$,Fe$_{0.06}$)As layers [12], and revealed a two-fold magnetic anisotropy along the [-110] direction and an 8-fold symmetric anisotropy along the crystal axes of (In,Fe)As, thus supporting the intrinsic ferromagnetism of this material.

The Arrott plots (MCD$^2$ vs. $H$/MCD) of sample A4 and B4 at different temperatures clearly show that sample A4 is ferromagnetic at $T < T_C$ ~ 34 K, and that of sample B4 is ferromagnetic at $T < T_{C-1}$ ~ 28 K (see Supplementary Information). Next, we confirm these by SQUID measurements. Figures 3(a) and (b) shows the field cooling (FC) and zero-field cooling (ZFC) magnetization ($M$) data of sample A4 and B4, measured by SQUID. A small magnetic filed of 20 Oe is applied in-plane along the GaAs[-110] direction. The $M$-$T$ curves of sample A4 show monotonic behavior both for FC and ZFC, which both rise at $T_C$ ~ 34 K. In contrast, sample B4 shows two-phase ferromagnetism; one is the matrix phase with $T_{C-1}$ ~ 28 K, and the other is the superparamagnetic phase with $T_{B-2}$ ~ 35 K and $T_{C2}$ ~ 70±10 K. Note that the MCD spectrum of sample B4 measured at 50 K (higher than $T_{C-1}$ and $T_{B-2}$ but lower than $T_{C2}$) still preserves clear features of the zinc-blende InAs structure (Fig. S2(c) in the Supplementary Information). This fact indicates that these clusters are *not* intermetallic precipitated particles but zinc-blende (In,Fe)As clusters with high concentration of Fe atoms.

Since the Fe concentration in sample B4 is as high as 8% (close to sample B0), zinc-blende clusters with high concentration of Fe atoms are formed due to the well-known spinodal decomposition phenomenon, consistent with the EDX data in Fig. 1(b). The *M-H* curves measured with a magnetic field applied along the [-110] direction in the film plane are shown in the inset of Figs. 3(a) and (b) for sample A4 and B4, respectively. The magnified in-plane *M – H* curves near the origins (bottom-right of the insets of Figs. 3(a) and (b)) show hysteresis but small remanent magnetization. Here, we propose a discrete multi-domain model (Fig. 3(c)) to explain such behavior of *M-H* curves. In our model, the (In,Fe)As layer contains separate macroscopic ferromagnetic domains with sizes much larger than the film thickness of 100 nm. These ferromagnetic domains have no interaction between each other (except for dipolar interactions), because they are separated by paramagnetic areas in-between. Thus, they have different orientations of magnetization at zero magnetic field, resulting in small total remanent magnetization. The existence of paramagnetic areas between these domains are suggested by the fact that the average effective magnetic moments at saturation ($m_{\text{eff}}$ = 2.2 and 1.7 $\mu_B$ per doped Fe atom for sample A4 and B4, respectively) are smaller than the expected $m_{\text{Fe}}$ = 5 $\mu_B$ for $Fe^{3+}$. We then used a magneto-optical (MO) imaging technique to visualize such ferromagnetic domains. Figure 3(d) shows the MO image of a local area (size 36 μm×36 μm) of sample B4, captured using an indicator garnet film in close contact with the surface of sample B4. The colored contrast in this image reflects the Faraday rotation of light going through the magnetic domains of the indicator garnet film oriented by the stray fields from the underlying (In,Fe)As layer [13]. Many magnetic dipoles (shown by yellow arrows) with different orientations were observed, revealing the underlying ferromagnetic domains of (In,Fe)As with sizes of ~ 10 μm. There are also areas with nearly zero Faraday rotations (white areas) between these magnetic dipoles, which can be attributed to the paramagnetic areas. The formation of these discrete multi-domain structures can be explained by the electron-induced ferromagnetism of (In,Fe)As and the phase separation due to the different concentration of Be double donors; areas with $n > 10^{19}$ cm$^{-3}$ (pink areas in Fig. 3(c)) are ferromagnetic but areas with $n < 10^{19}$ cm$^{-3}$ (white) are not. The calculated shape anisotropy ($1.0 \times 10^3$ kJ/m$^3$) of sample B4 based on our discrete multi-domain model is consistent with the measured value ($1.2 \times 10^3$ kJ/m$^3$) (see Supplementary Information).

In Figs. 4(a) and (b), we show the evolution of ferromagnetism expressed by $T_C$ vs. electron concentration and resistivity vs. temperature of series A and B samples, respectively. It is clear that there is a threshold electron concentration of about $1 \times 10^{19}$ cm$^{-3}$ for (In,Fe)As to become ferromagnetic. The steep change in magnetic behavior at $1 \times 10^{19}$ cm$^{-3}$ shown in Fig. 4(a) is clearly correlated with the metal-insulator transition of (In,Fe)As layers as shown in Fig. 4(b). All of these results confirm that (In,Fe)As is an intrinsic n-type ferromagnetic semiconductor whose ferromagnetism is induced by electrons. It should be noted that sample A4 with $T_C$ as high as 34 K requires only an electron concentration of $1.8 \times 10^{19}$ cm$^{-3}$.

Comparing with (In,Mn)As, this electron concentration is an order of magnitude smaller ($T_C \sim$ 20 K requires $1.0 - 1.6 \times 10^{20}$ cm$^{-3}$ of holes for (In,Mn)As [14]).

In conclusion, we have grown an Fe-based n-type electron-induced FMS, (In,Fe)As. MCD, SQUID, and magnetotransport data show clear evolution of ferromagnetism in (In,Fe)As when increasing the electron density by chemical doping with a fixed Fe concentration. Development of such Fe-based n-type electron-induced FMS will open the way to fabricate all-FMS spintronic devices, as well as help understanding the physics of carrier-induced ferromagnetisms in FMS.

**Acknowledgements** This work was supported by Grant-in-Aids for Scientific Research including the Specially Promoted Research, the Special Coordination Programs for Promoting Science and Technology, the FIRST Program of JSPS, and the Global COE program (C04). The authors acknowledge Dr. Fujii for his help in the measurement of Seebeck effect. Part of this work was done at the Cryogenic Center, the Univ. Tokyo.

**Figure Captions**

**Fig. 1. (color) (a)** Transmission electron microscopy (TEM) image of a 100 nm-thick $(In_{0.909},Fe_{0.091})As$ layer (sample B0 in table I) grown on a GaAs substrate, taken from the GaAs[110] direction. **(b)** In, Fe and As atomic concentrations obtained by energy dispersive X-ray spectroscopy taken at 6 points marked by * in the above TEM image. **(c)** High-resolution TEM lattice image taken at an $(In_{0.909},Fe_{0.091})As$ area close to the substrate (marked by the red rectangle in Fig. 1a). Inset shows the transmission electron diffraction of this (In,Fe)As layer. **(d)** Three-dimensional atom distribution of Fe, In, and As in a 100 nm-thick $(In_{0.95},Fe_{0.05})As$ layer (sample A4), obtained by the laser assisted three-dimensional atom probe technique. One dot (red, green, blue) corresponds to one (Fe, In, As) atom.

**Fig. 2. (color online)** Magnetic circular dichroism spectra (MCD) of **(a)-(d),** $(In_{0.95},Fe_{0.05})As$ samples (A1 - A4 in table I) with electron concentrations of $1.8 \times 10^{18}$,

$2.9\times10^{18}$, $6.2\times10^{18}$, $1.8\times10^{19}$ cm$^{-3}$, respectively, and **(e)-(h), (In$_{0.92}$,Fe$_{0.08}$)As** samples (B1 - B4 in table I) with electron concentrations of $1.3\times10^{18}$, $1.5\times10^{18}$, $9.4\times10^{18}$, $2.8\times10^{19}$ cm$^{-3}$, respectively, measured at 10 K and under a magnetic field of 1 Tesla applied perpendicular to the film plane. With increasing the electron and Fe concentrations, the MCD spectra show strong enhancement at optical critical point energies $E_1$ (2.61 eV), $E_1 + \Delta_1$ (2.88 eV), $E_0$' (4.39 eV) and $E_2$ (4.74 eV) of InAs. **(i)** MCD spectrum of a 44 nm-thick Fe thin film grown on a GaAs substrate at 30°C. The spectrum is clearly different from that of (In,Fe)As.

**Fig. 3. (color) (a)(b)** Magnetization ($M$ - $T$ curves) of sample A4 and B4, measured under 1-Tesla field-cooling (FC) and zero-field-cooling (ZFC) conditions. The magnetic field (20 Oe) is applied in-plane along the GaAs[-110] direction. The insets show the magnetization hysteresis loops ($M$ - $H$) of sample A4 and B4 measured at 10 K. The magnified $M$ - $H$ curves near the origin are shown in the bottom-right of the insets, which show the remanent magnetization and coercive forces. **(c)** Discrete multi-domain model (plan view of the (In,Fe)As film). Pink areas (indicated by dashed lines) are ferromagnetic with $n > 10^{19}$ cm$^{-3}$, while white areas with $n < 10^{19}$ cm$^{-3}$ are not. Magnetization directions of ferromagnetic domains are indicated by yellow arrows. Each domain has size much larger than the film thickness of 100 nm (see text). **(d)** Magneto-optical imaging of sample B4 under zero magnetic field at 4 K. The light source is a halogen lamp with a white light. Discrete

ferromagnetic domains (shown by green dotted circles) with sizes of ~ 10 μm are visible. Areas with small Faraday rotation (white) between ferromagnetic domains correspond to paramagnetic areas.

**Fig. 4. (color online) (a)** $T_C$ vs. electron concentration and **(b)** resistivity vs. temperature summarized for sample series A and B. An electron concentration threshold of about $10^{19}$ cm$^{-3}$ is needed for ferromagnetism, which is also the boundary for metal-insulator transition.

**Table I.** List of $(In_{1-x},Fe_x)As$ samples. All samples were grown at 236°C.

| Sample | $x$ (%) | Electron concentration $n$ (cm$^{-3}$) | Non magnetic dopants |
|---|---|---|---|
| A1 | 5.0 | $1.8 \times 10^{18}$ | Be |
| A2 | 5.0 | $2.9 \times 10^{18}$ | Be |
| A3 | 5.0 | $6.2 \times 10^{18}$ | Be |
| A4 | 5.0 | $1.8 \times 10^{19}$ | Be |
| B0 | 9.1 | $1.6 \times 10^{18}$ | None |
| B1 | 8.0 | $1.3 \times 10^{18}$ | Be |
| B2 | 8.0 | $1.5 \times 10^{18}$ | Be |
| B3 | 8.0 | $9.4 \times 10^{18}$ | Be |
| B4 | 8.0 | $2.8 \times 10^{19}$ | Be |

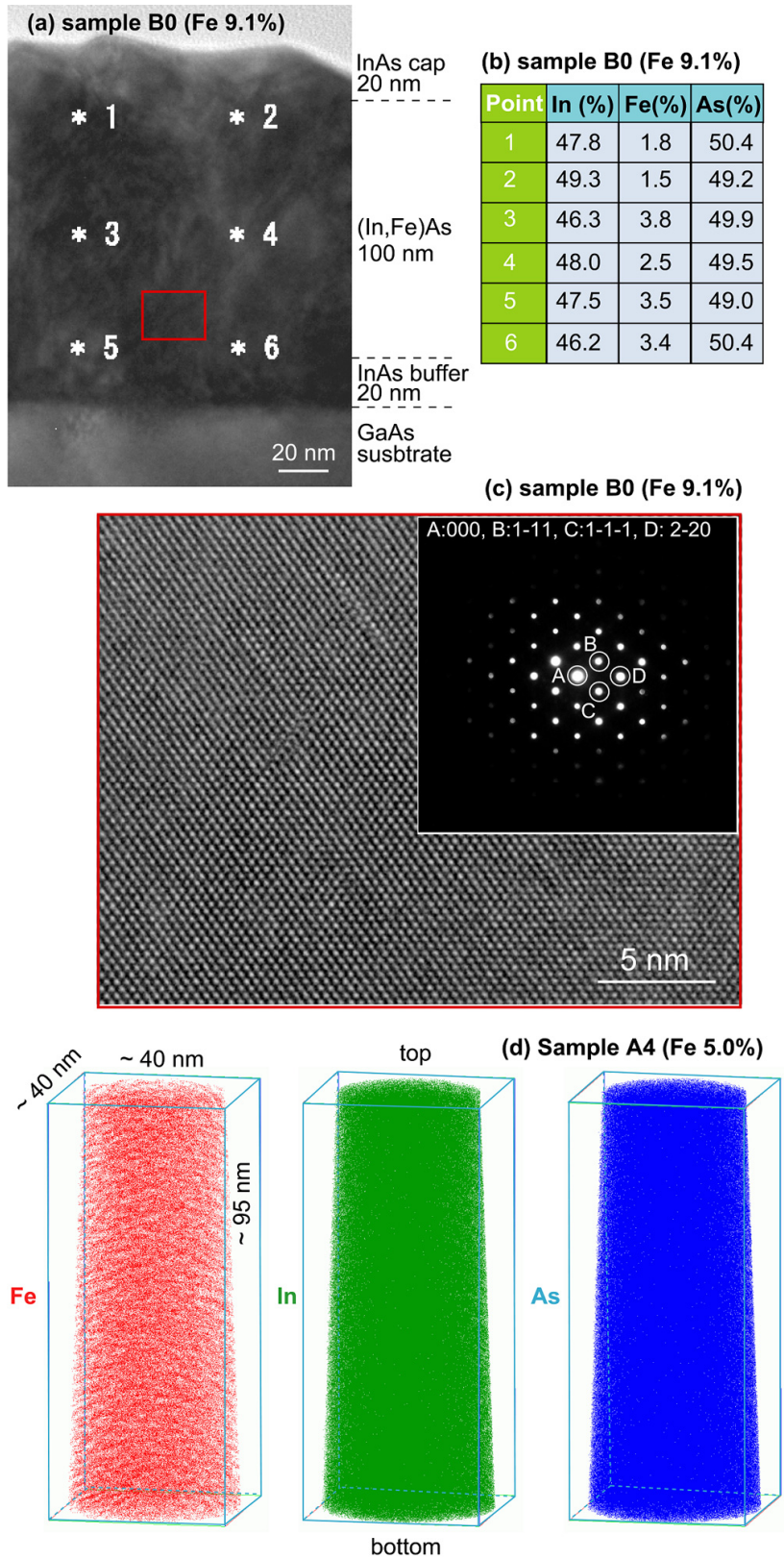

Fig. 1. Hai *et al.*

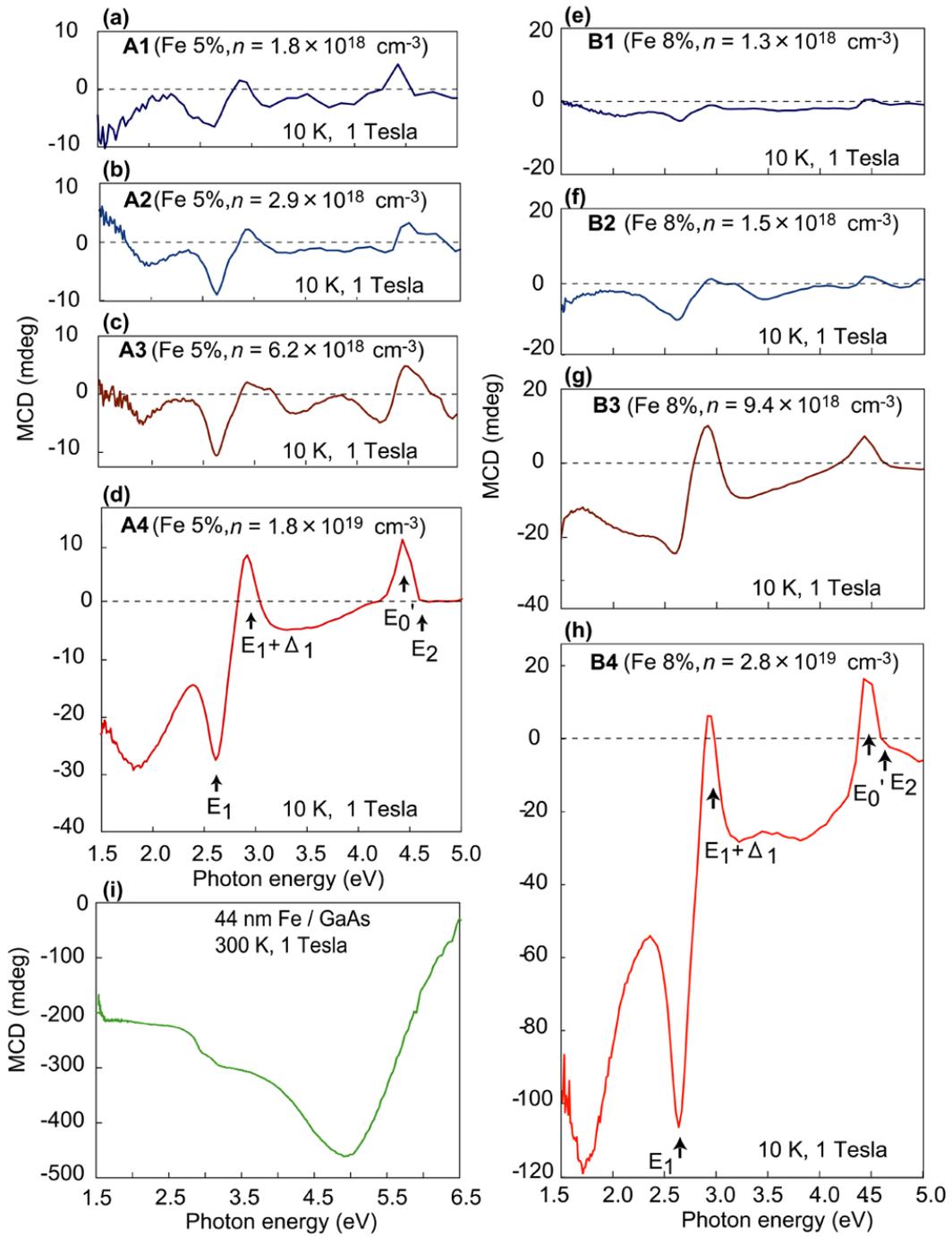

Fig. 2. Hai *et al.*

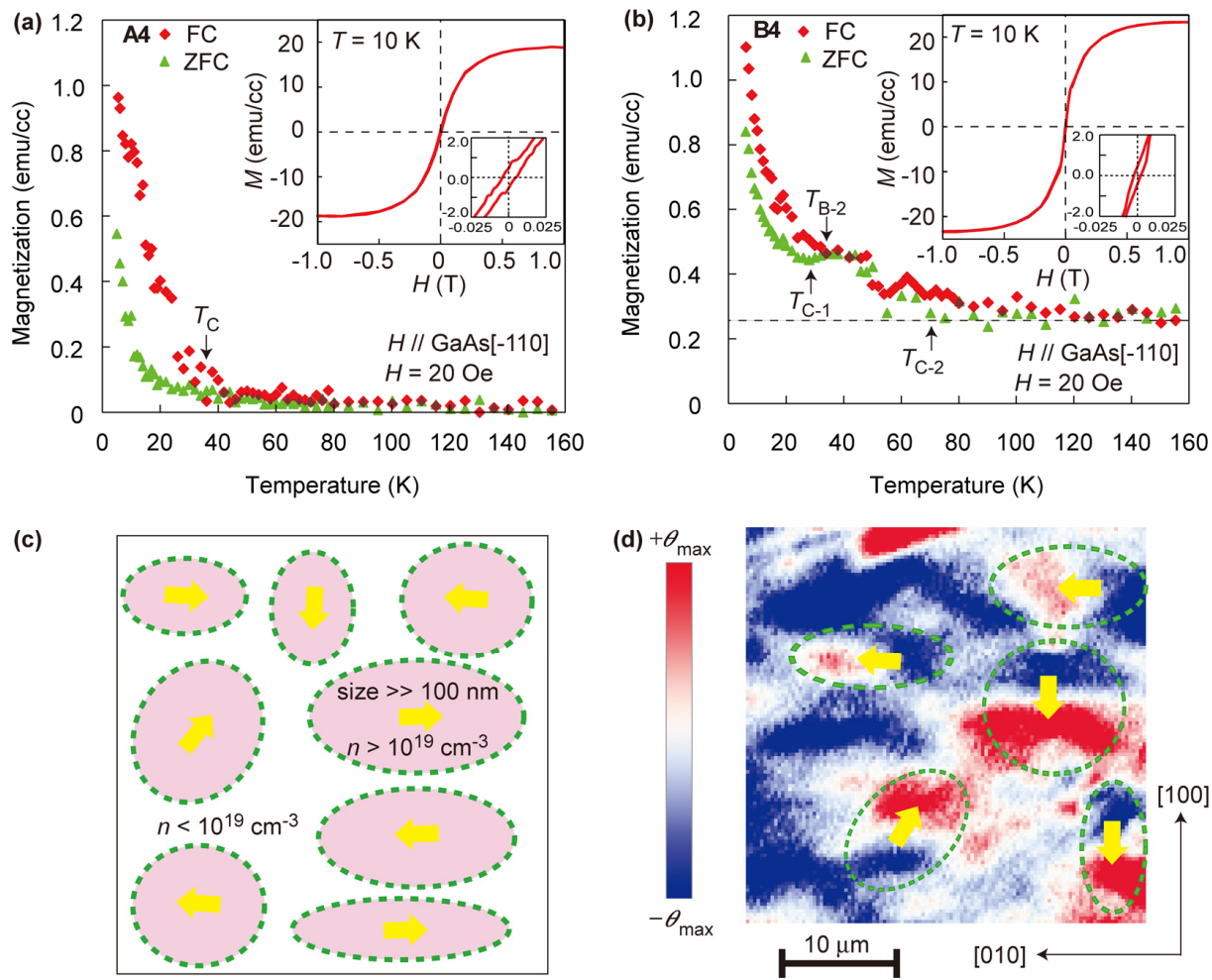

Fig. 3. Hai *et al.*

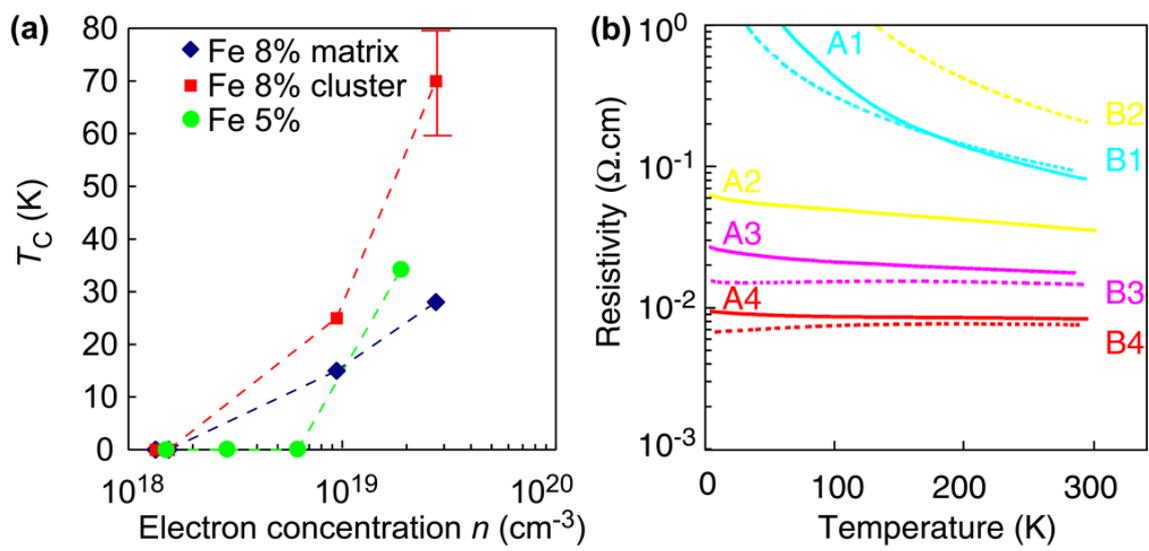

Fig. 4. Hai *et al.*

# Supplementary Information

# Fabrication and characterization of n-type electron-induced ferromagnetic semiconductor (In,Fe)As

Pham Nam Hai, Le Duc Anh, Shyam Mohan, Tsuyoshi Tamegai, Masaya Kodzuka,

Tadakatsu Ohkubo, Kazuhiro Hono and Masaaki Tanaka.

**Temperature dependence of the mobility of sample B0**

Fig. S1 shows the temperature dependence of the electron mobility $\mu$ of the as-grown $(In_{0.909},Fe_{0.091})As$ (sample B0), with vertical and horizontal axes plotted in the logarithmic scale. The dashed red line is the fitting $\mu \sim T^{\gamma}$. For $T < 50$ K, the mobility is nearly temperature-independent ($\gamma = 0.04$). For $T > 50$ K, the mobility still weakly depends on temperature ($\gamma = 0.25$). These suggest that the Fe impurities in this material remain neutral. If the Fe impurities were ionized (i.e. in the acceptor $Fe^{2+}$ state), the sample would be p-type and the $\mu - T$ relation would be given by $\mu \sim \frac{1}{n_{ion}}(2k_BT)^{3/2}[\ln(1+\alpha k_B^2 T^2)]$ for ionized impurity scattering, requiring $\gamma \geq 1.5$. In reality, the sample is n-type and $\gamma$ is close to zero, indicating that the Fe impurities remain in the neutral state. When the neutral impurity scattering dominates, the $\mu - T$ relation is given by $\mu \sim \frac{1}{n_{neutral}} \times const$, which is nearly temperature-independent. (Here, we neglect the contributions from phonon scattering, which are confirmed to be negligible by theoretical calculations. The contribution from alloy scattering can not be calculated quantitatively. However, it can not be a dominant scattering mechanism since it would give a negative $\gamma$.) Therefore, the Fe impurities on In sites should be in the $Fe^{3+}$ state. This result is similar to that obtained for paramagnetic (Ga,Fe)As [1]. Since the Fe impurities contribute to spin (magnetisation) but not to carrier generation, we have an important degree of freedom for controlling the carrier type and carrier concentration by independent chemical doping.

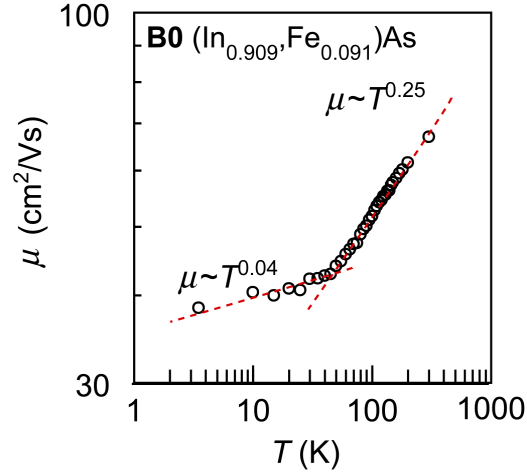

**Fig. S1.** Temperature dependence of the electron mobility of sample B0, which indicates the neutral state of Fe impurities on In sites. Dashed red line is the fitting $\mu \sim T^{\gamma}$.

**Normalized MCD spectra**

Figures S2a-c show the normalized MCD spectra of sample A4 and B4, measured at 0.2, 0.5 and 1 Tesla. In Fig. S2a, the normalized spectra of sample A4 show nearly perfect overlapping on a single spectrum over the whole photon-energy range, proving that the MCD spectra comes from a single phase ferromagnetism of the whole (In,Fe)As film. In Fig. S2b, the normalized spectra of sample B4 shows perfect overlapping in the range of 2.5 – 5 eV, but deviate slightly from a single spectrum at photon energies lower than 2.5 eV. The peak at 1.8 eV develops faster than that at $E_1$ at low magnetic field, but they approach each other at 1 Tesla. This shows that the MCD spectra of sample B4 come from two ferromagnetic phases; one is the (In,Fe)As matrix phase having a MCD spectrum similar to that of sample A4 (Fig. S2a), and the other is the cluster phase whose spectrum is enhanced at low photon energy (< 2.0 eV) as shown in Fig. S2c. The latter turned out to be superparamagnetic zinc-blende (In,Fe)As clusters with higher density of Fe, as clarified by SQUID measurements.

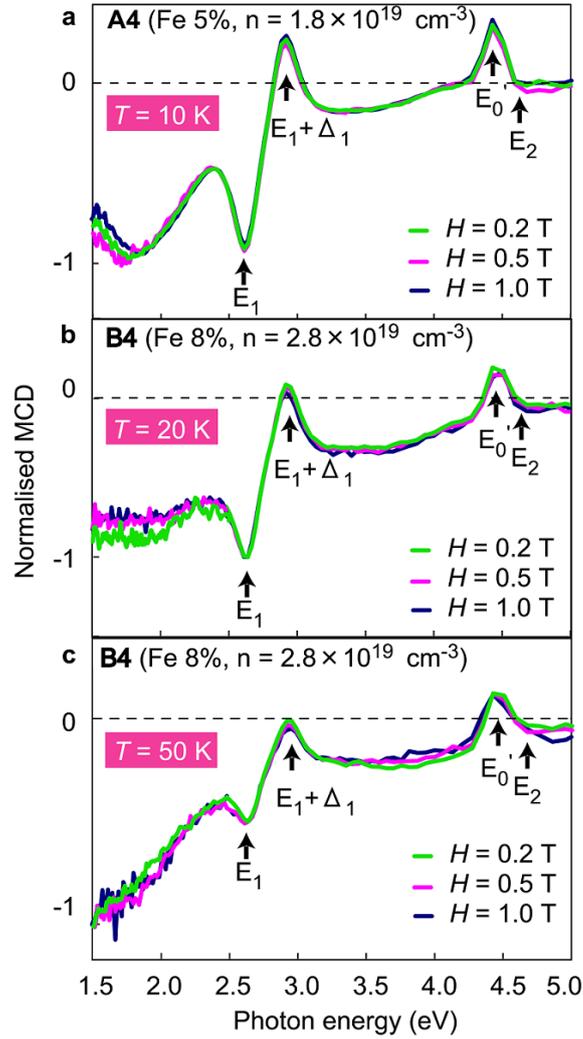

**Fig. S2.** Normalised MCD spectra of **a,** sample A4 with $H$ = 0.2, 0.5 and 1 Tesla, measured at 10 K. A single spectrum over the whole photon-energy range proves the single phase ferromagnetism of this sample, and **b-c,** sample B4 with $H$ = 0.2, 0.5 and 1 Tesla, measured at 20 and 50 K, respectively. The spectra at 20 K shows two phase ferromagnetism at low photon energy (< 2.0 eV), and can be decomposed to two components. One is the matrix with spectrum similar to the sample A4 (Fig. S2a), and the other is the cluster phase whose spectrum is enhanced at low photon energy (< 2.0 eV). At 50 K when there is only the super paramagnetic cluster phase, the MCD spectrum still preserves the zinc-blende band structure, revealing that those clusters are zinc-blende (In,Fe)As with high Fe concentration due to spinodal decomposition.

# Arrott plots (MCD² vs. *H*/MCD) of sample A4 and B4

Figures S3a and S3b shows the MCD-*H* curves of sample A4 and B4, respectively, measured at different temperatures and at the photon energy of 2.6 eV. From these data we can deduce the Curie temperature of these samples by the Arrott plot technique. Figures S3c and S3d show the Arrott plots MCD² vs. *H*/MCD of sample A4 and B4 at different temperatures, where MCD is the MCD intensity which is proportional to the magnetisation *M*. It is clear that sample A4 is ferromagnetic at $T < T_C \sim 34$ K, and the matrix of sample B4 is ferromagnetic at $T < T_{C-1} \sim 28$ K.

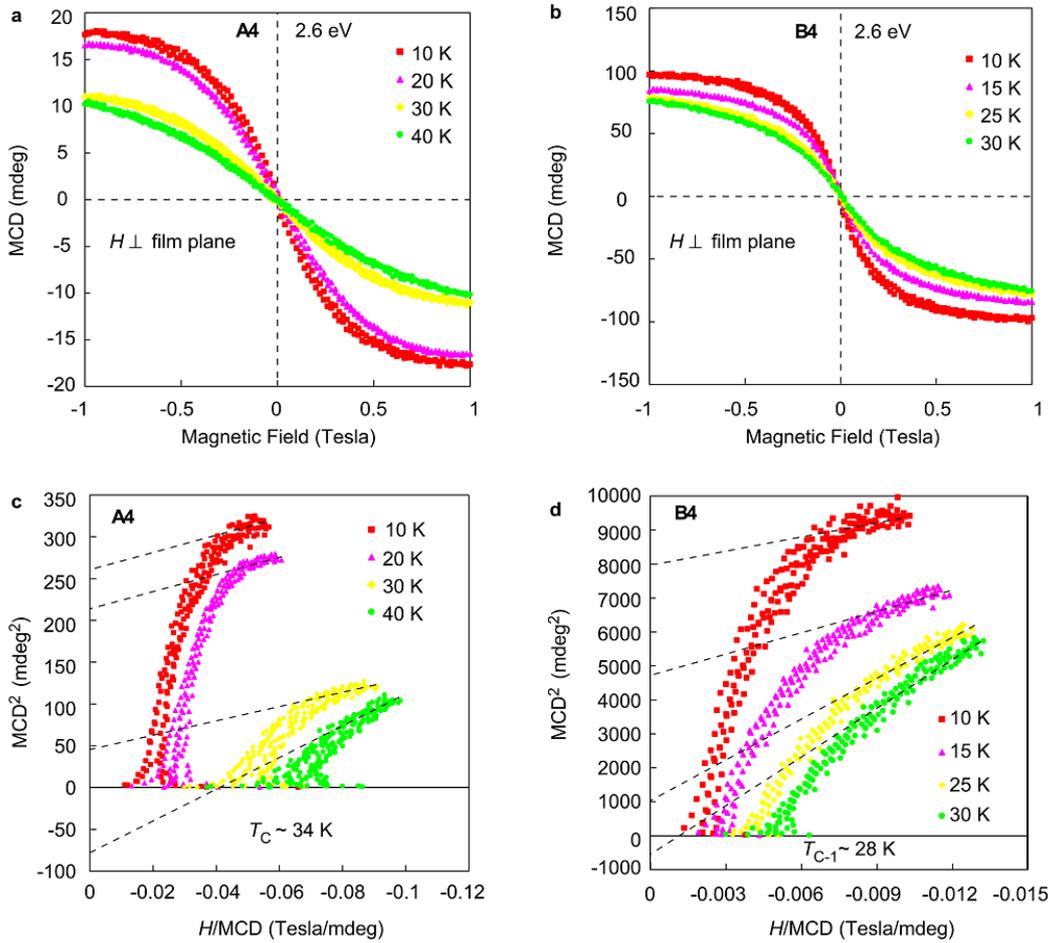

**Fig. S3. a,b,** MCD-*H* curves at different temperatures of sample A4 and B4, respectively. These curves are measured at the photon energy of 2.6 eV. **c,d,** Arrott plot (MCD² vs. *H*/MCD) of sample A4 and B4. Sample A4 is ferromagnetic at $T < T_C \sim 34$ K. The matrix of sample B4 is ferromagnetic at $T < T_{C-1} \sim 28$ K.

**Paramagnetic MCD-*H* characteristics of sample A4 and B4 at *T* > *T*$_C$**

Figures S4a and S4b show the MCD-*H* characteristics at temperatures higher than the Curie temperature $T_C$ = 34 K for sample A4 (Fe 5%, $n$ = 1.8×10$^{19}$ cm$^{-3}$), and the Curie temperatures $T_{C-1}$= 28 K and $T_{C-2}$= 70 K ± 10 K for sample B4 (Fe 8%, $n$ = 2.8×10$^{19}$ cm$^{-3}$). It is seen that the MCD-*H* curves show paramagnetism in these high temperature ranges. Note that there is also no super-paramagnetism, whose Langevin-function shape is expected at high temperature but there is no such shape.

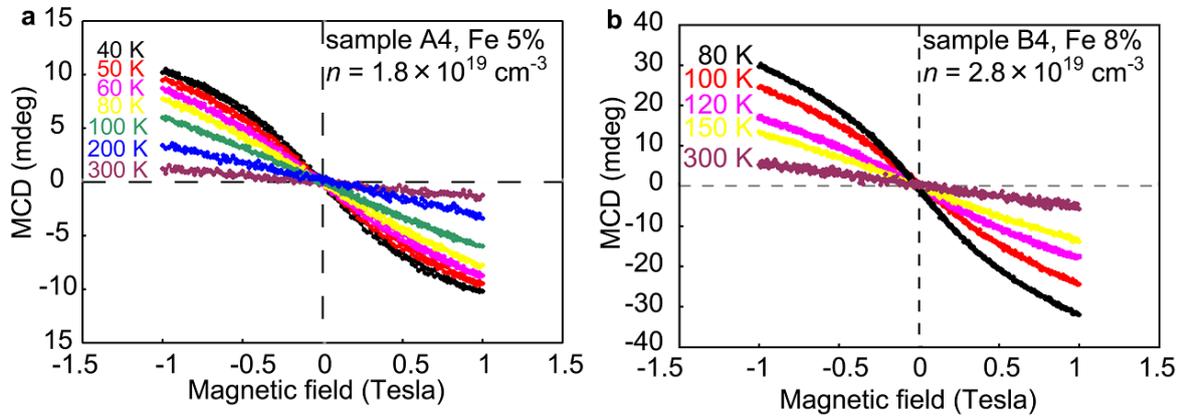

**Fig. S4**. Paramagnetic MCD-*H* characteristics of **a,** sample A4 (Fe 5%, $n$ = 1.8×10$^{19}$ cm$^{-3}$, $T_C$ = 34 K) at 40 K – 300 K, and **b,** sample B4 (Fe 8%, $n$ = 1.8×10$^{19}$ cm$^{-3}$, $T_{C-1}$ = 28 K, $T_{C-2}$ = 70±10 K) at 80 K – 300 K. All are measured at photon energy of 2.6 eV.

**Transport characteristics.**

Figures S5a and S5b show the Hall resistance of sample A4 and B4, respectively. The normal Hall effect with negative gradient, showing the n-type conduction of these (In,Fe)As layers, dominates the Hall voltage. The n-type conductivity is also confirmed by the polarity of the thermoelectric Seebeck coefficient (see Fig. S6). There is a small fraction (~ 3%) of *positive* anomalous Hall effect (AHE) contribution in both samples due to spin-dependent scattering of electrons at Fe sites, as shown in the anomalous Hall resistance (AHR) curves in Fig. S5c and S5d. The weak AHE in n-type FMS compared with that of p-type FMS is indeed consistent with Berry-phase theory of AHE in FMS [2].

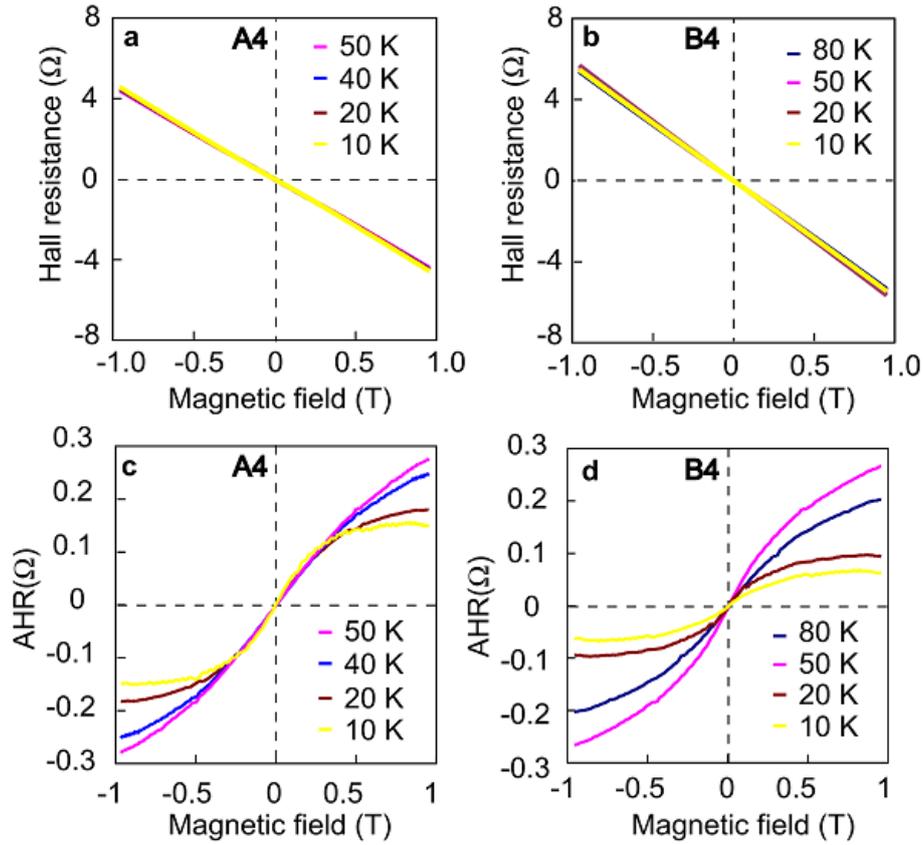

**Fig. S5. a-b,** Hall resistances of sample A4 and B4, respectively. The Hall resistance is dominated by the normal Hall effect with negative gradient, showing the n-type conduction of these samples. **c-d,** The extracted *positive* anomalous Hall resistances (AHR) for sample A4 and B4, respectively. The AHR are about 3% of the normal Hall resistances, even at 10 K.

**Thermoelectric Seebeck effect at room temperature**

A convenient way to confirm the carrier type in heavily doped semiconductors is the thermoelectric Seebeck effect. Figure S6a shows the principle of the Seebeck effect. When there is a temperature gradient $\Delta T$ between two edges of a sample, carriers at the hot side are more thermally activated and then diffuse to the cold side until equilibrium is established. As a result, a voltage $\Delta V$ will be generated between the edges. The Seebeck coefficient $\alpha$ of a material is defined as $\alpha = -\Delta V/\Delta T$. If carriers are electrons, $\alpha$ is negative. Inversely, if carriers are holes, $\alpha$ is positive. Figure S6b shows the experimental setup to measure the Seebeck effect of our (In,Fe)As at room temperature. The hot side is a copper (Cu) electrode with a heater, placed on an epoxy film. The epoxy film acts as a thermal insulator. The cold side is a Cu electrode placed on a sapphire substrate, which acts as a thermal sink. A piece of sample bridges the hot and cold electrodes. Silver paste is used for electrical contacts between the edges of the sample and the electrodes. Voltage signals from a thermocouple made from Cu wire (red thin line) and Constantan (green thin line) measure the temperature difference $\Delta T_{raw}$ between the hot Cu electrode and the sapphire substrate when the heater is turned on. Figure S6c shows the measured $\Delta V$ - $\Delta T_{raw}$ of sample B4 at different heater currents. It is clear that $\alpha$ is negative from the gradient of this data. Thus, the carriers are electrons, which is consistent with the Hall effect measurement results described in the main text. All other samples mentioned in the paper have the same result, i.e. n-type conduction.

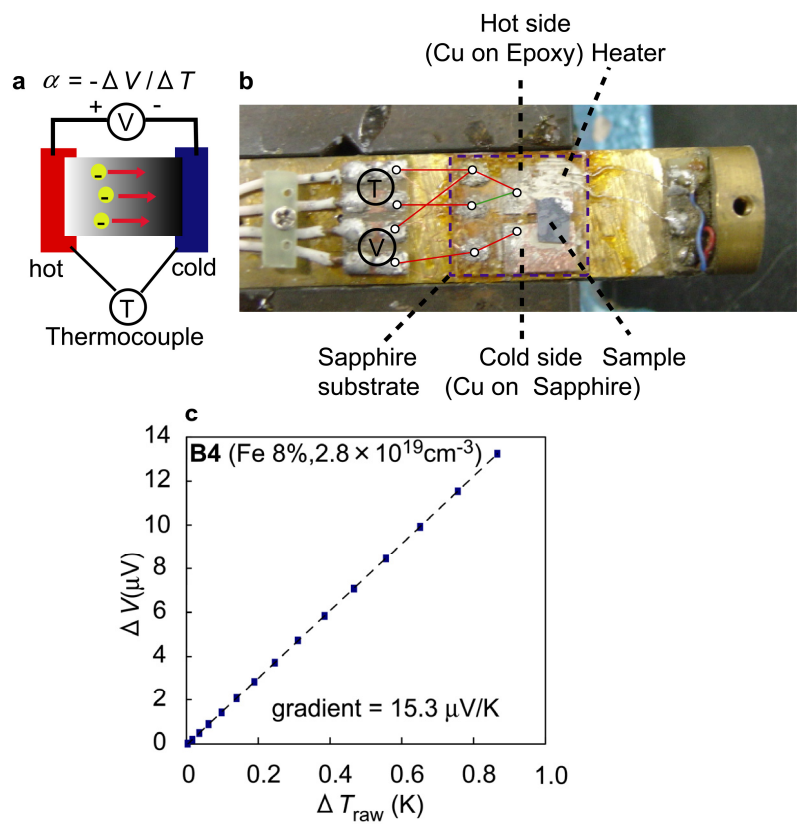

**Fig. S6. a,** Principle of thermoelectric Seebeck effect. **b,** Experiment setup to measure the Seebeck effect. **c,** Measured $\Delta V$ - $\Delta T_{raw}$ of sample B4.

**Electron concentration vs. doped Be concentration**

Figure S7 shows the electron concentration $n$ vs. doped Be concentration $n_{Be}$ for sample series A and B. $n$ are measured at room temperature. When doped at low growth temperature (236°C), Be atoms become donors rather than acceptors, whereas Be atoms become acceptors when doped at a normal growth temperature of InAs (400°C). Note that similar donor behavior of group II dopants in InAs grown at low substrate temperature has been observed for the case of Mn in (In,Mn)As [3]. The relationship between $n_{Be}$ vs. $n$ is not trivial as shown in Fig. S7. The highest $n$ is obtained when $n_{Be}$ is around $10^{19}$ cm$^{-3}$. At this doping level, the electron concentration is twice as large as the Be concentration. This suggests that Be atoms reside at interstitial positions and act as double donors. The double-donor behavior of interstitial Be has also been observed when doped in silicon carbide [4]. However, when increasing $n_{Be}$ up to $10^{20}$ cm$^{-3}$, $n$ decreased to $10^{18}$ cm$^{-3}$. Be doping levels except for $n_{Be} \sim 1\times10^{19}$ cm$^{-3}$ resulted in $n < n_{Be}$. The decrease of electron concentration at high $n_{Be}$ may be due to compensation effects due to crystal defects or Be atoms that partly enter the substitutional sites.

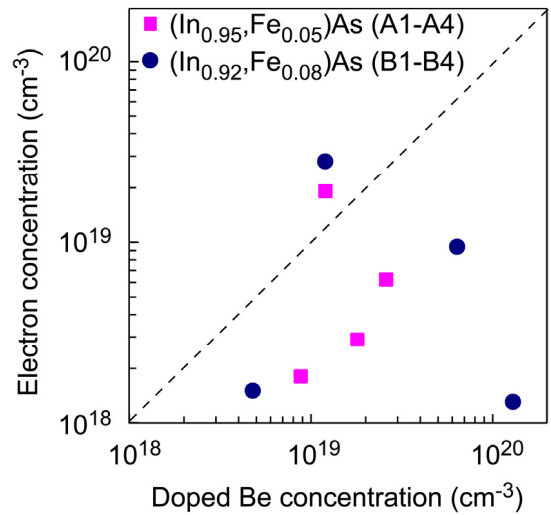

**Fig. S7.** Electron concentration vs. doped Be concentration.

**Shape magnetic anisotropy of (In,Fe)As**

In this section, we will show that the measured shape magnetic anisotropy of (In,Fe)As can be well explained by the discrete multi-domain model. Figure S7 shows the *M-H* curve of sample B4 measured with *H* applied in the film plane and perpendicular to the plane. The *M-H* curve with *H* applied in the film plane saturates at lower magnetic field than that with *H* applied perpendicular to the plane. The difference appearing for $H > 0.1$ Tesla comes from the shape anisotropy of the macroscopic ferromagnetic domains whose sizes are much larger than the film thickness of 100 nm. The shape anisotropy energy is measured to be 1.2 kJ/m$^3$. The shape anisotropy of discrete multi-domains is given by $C*\mu_0 M_S^2/2$, where $\mu_0 M_S^2/2$ is the shape anisotropy of a single macroscopic domain (size >> 100 nm), and *C* = (volume of ferromagnetic domains / total volume) is the volume filling factor of ferromagnetic domains. Using calculated saturation magnetisation $M_S = 67$ emu/cc assuming $m_{Fe} = 5\mu_B$ and $C = (1.7\mu_B)/(5\mu_B) = 0.34$, we obtain $C*\mu_0 M_S^2/2 = 1.0$ kJ/m$^3$ for sample B4, which is in good agreement with the experimental value.

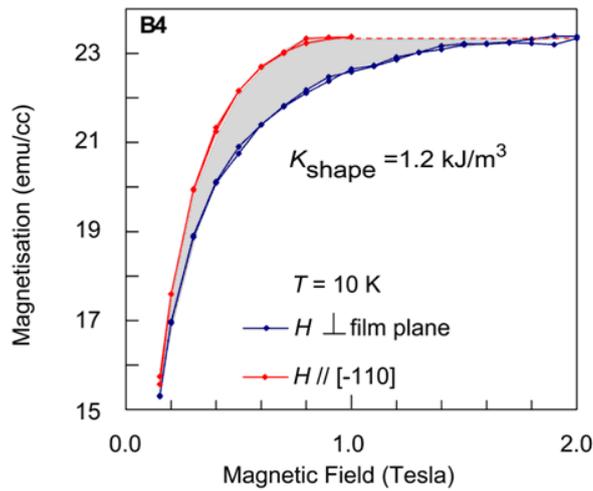

**Fig. S7.** *M-H* curves for $H > 0.1$ Tesla of sample B4 measured with a magnetic field *H* applied in the film plane (red) and perpendicular to the film plane (blue).

**Magneto-optical imaging: field dependence and temperature dependence**

Figures S8a and b show the field and temperature dependence of magneto-optical (MO) images of a local area (size 36 μm×36 μm) of sample B4, respectively. In Fig. S8a, the magnetic field was applied perpendicular to the film plane. The field was swept from +400 Oe to -400 Oe with a step of 100 Oe. Then, the field was reduced from -400 Oe to 0 Oe. In this local area, the MO contrast is reversed between $H$ = -100 Oe (image F) and $H$ = -200 Oe (image G), indicating a coercive force larger than 100 Oe. It should be noted that the remanent MO contrast of image E is perfectly reversed to that of image L, confirming the existence of ferromagnetic multi-domains in (In,Fe)As. Figure S8b shows the temperature dependence of the MO contrast in image L. The contrast becomes weaker as temperature increases, and there is no contrast at $T$ = 150 K, which is higher than the Curie temperatures of sample A4.

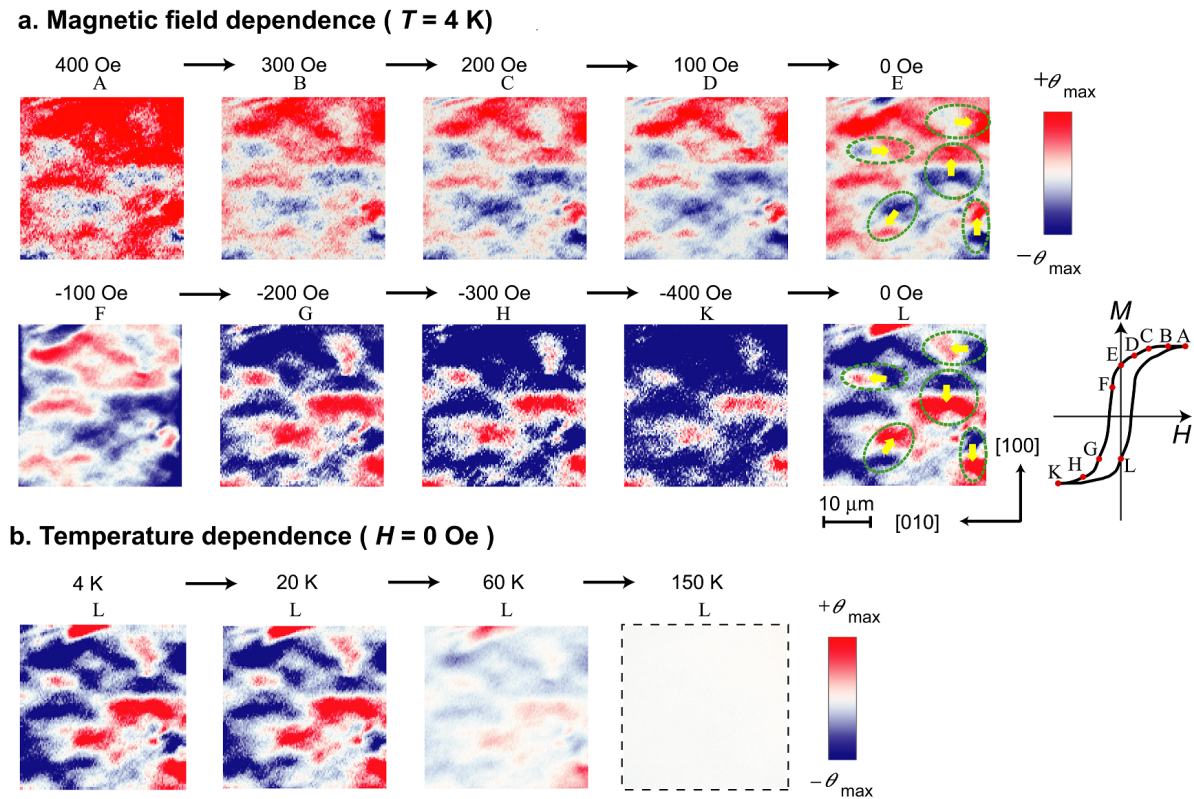

**Fig. S8.** a, Magnetic field dependence, and b, temperature dependence of MO images of a local area (size 36 μm×36 μm) of sample B4.